\documentclass[twocolumn,final,balancelastpage,superscriptaddress]{revtex4}

\usepackage{graphicx}

\usepackage{tabularx}
\usepackage{amssymb}
\usepackage{amsmath}
\usepackage{amsbsy}
\usepackage{comment}
\usepackage{mathrsfs}
\usepackage{dsfont}
\usepackage{nicefrac}
\usepackage{wrapfig}
\usepackage{amsfonts}
\usepackage{url} % link package
\usepackage[nolist]{acronym}
\usepackage{lineno}%(enumera las lineas)
\usepackage{nicefrac}
\usepackage{subfigure}
\usepackage{epsf,color,colordvi,pifont}
\usepackage{showkeys}
\usepackage{esint}

\DeclareFontFamily{OT1}{pzc}{}
\DeclareFontShape{OT1}{pzc}{m}{it}{<-> s * [1.10] pzcmi7t}{}
\DeclareMathAlphabet{\mathpzc}{OT1}{pzc}{m}{it}

\begin{document}
\title{Condensation of slow  $\gamma$-quanta in strong magnetic fields}
\author{L. \surname{Folkerts}}
\affiliation{Institut f\"{u}r Theoretische Physik, Heinrich-Heine-Universit\"{a}t D\"{u}sseldorf,\\ Universit\"{a}tsstr.\,1, 40225 D\"{u}sseldorf, Germany}
%\email{Leah.Folkerts@hhu.de}
\author{R. \surname{Egger}}
%\email{egger@hhu.de}
\author{C. \surname{M\"{u}ller}}
%\email{c.mueller@tp1.uni-duesseldorf.de }
\author{S.  \surname{Villalba-Ch\'avez}}
\email{villalba@uni-duesseldorf.de}
\affiliation{Institut f\"{u}r Theoretische Physik, Heinrich-Heine-Universit\"{a}t D\"{u}sseldorf,\\ Universit\"{a}tsstr.\,1, 40225 D\"{u}sseldorf, Germany}

\begin{abstract}
The implications of the root singularity of the vacuum polarization tensor near the first pair creation threshold on blackbody radiation are investigated for magnetic fields above the characteristic scale of quantum electrodynamics. We show  that the vacuum birefringence in such a strong background leads to an anisotropic behavior of the Planck radiation law. The  thermal spectrum is characterized by a resonance that competes with the Wien maximum, causing a crossover in the  low $\gamma$-spectrum  of the  heat radiation. A light state resembling a  many-body condensate with slow  motion is linked to  the  high-temperature phase. This novel state of radiation  may coexist with nuclear or quark  matter  in a neutron star's core, increasing its compactness and influencing its stability.
\end{abstract}

\date{\today}

\maketitle

{\it Introduction}---Considerable experimental progress has been made in the field of photonic Bose-Einstein Condensates (BECs) over the past decade, attaining this coherent macroscopic phase in a variety of systems driven by pumped dye-filled microcavities \cite{KlaersNatPhys,Klaers,KlaersNatComm}, doped fibers \cite{Weill1,Weill2}, and semiconductor quantum well microresonators \cite{Barland,Schofield}. In vacuum, however, a photon ensemble cannot sustain a BEC, despite obeying Bose-Einstein statistics, because photons are massless, which limits their existence to states above the lowest energy.  Moreover, there is no thermalization mechanism to sustain a constant number of quanta as the system cools down.  Condensing photons into any other state is \emph{a priori} plausible if quantum vacuum properties---such as Lorentz invariance---are altered  by the presence of an external magnetic field $\pmb{B}$. Polarization of quantum vacuum fluctuations then renders vacuum akin to a dielectric material.  Modifications of this type---accounted for by the polarization tensor $\Pi_{\alpha\beta}$ \cite{Batalin,Wu,Baier,Urrutia,Hattori1,Hattori2,Felix}---are expected to become relevant at field strengths $B\gtrsim B_{\mathrm{cr}}=\frac{m^2}{\vert e\vert}=4.4\times 10^{13}\; \rm G$,   as the  photon energy approaches any of the $\Pi_{\alpha\beta}$-singularities \cite{ShabadLett,Shabad,Baier}  at the creation thresholds of  electron-positron pairs, with each lepton occupying a  Landau level. This cyclotron resonance  is  linked to the nonlocal feature of quantum electrodynamics by which  the photon dispersion laws  significantly deviate from the  light cone shape  \cite{ShabadLett,Shabad,Witte,ShabadJETP}, and  results from  the photon's coexistence with a quasi-formed pair,  resembling  the behavior  in semiconductors when the radiation frequency approaches a crystal's absorption line and an exciton-polariton is formed \cite{Ashcroft}.

Although the described phenomenology has been known for some time and has  direct implications for both the splitting  \cite{AdlerPRL1,Adler:1971wn,Baring,Baring:2000cr,Usov:2002df,Chistyakov:2012ms} and the photon capture effects \cite{ShabadNature,ShabadAstroSpaSci,Herold,Bathia,Selym2}, its consequences for thermal radiation have yet to be explored. Certainly, any deviation from the photon dispersion relation is expected to have non-trivial consequences in the Planck radiation law, unveiling new quantum states of light, disclosing novel phase transitions, and perhaps providing insight into the conditions needed for photons to condense.   These hitherto unexplored blackbody properties have immediate astrophysical implications for pulsars, where surface magnetic fields  $B\sim 10^{12}-10^{15}\;\rm G$  have been inferred \cite{Manchester,Kouveliotou,Alaa,Gnedin,Ibrahim,McGill,Kaspi}, and could  soon be probed at CERN via peripheral heavy-ion collisions, where fields $B\gtrsim 10^{15}\;\rm G$  are projected  to emerge    \cite{Kharzeev,Skokov,Zhong,Brandenburg}. Moreover, they gain cosmological relevance in view of the plausible existence of  even  stronger strengths  $B\sim 10^{24}\; \rm G$  during the electroweak phase transition \cite{Vashaspati,Enqvist,Baym,Grasso:2000wj,Vachaspati:2020blt}. These potential repercussions call  for a systematic analysis of the cyclotron resonance in vacuum as it pertains to blackbody radiation.  

In this letter, we show that the strong refraction caused by this phenomenon near the first pair creation threshold  leads to the emergence of a spike  in the blackbody spectrum which is suppressed for temperatures $T\ll m\ll \sqrt{\vert eB\vert}$, but  dominates over the Wien peak  when  $m\lesssim T\ll \sqrt{\vert eB\vert}$, causing a crossover in  the corresponding $\gamma$-radiation.  Thus, at low $T$, the phase approximately obeys the  standard Planck radiation law. In contrast, at high $T$, the phase is dominated  by a single photon species whose  group velocity   is considerably smaller than the speed of light $c$.  We show that the corresponding ensemble of slow $\gamma$ quanta exhibits the characteristics of a condensate, and further reveal that it may exist in neutron star (NS) cores, where it deepens the gravitational potential without providing significant support for hydrostatic equilibrium or affecting the stellar cooling.

{\it Theoretical Framework}---The theoretical framework for describing the inherent nonlocal nature of the cyclotron resonance in vacuum  will be based on the generating function of one-particle irreducible vertices of QED \cite{Fradkin,Holger}:
\begin{equation}
\begin{split}
 \Gamma&=-\frac{1}{4}\int d^4x\, f_{\alpha\beta}f^{\alpha\beta}\\
 &\quad+\frac{1}{2}\int d^4x \int d^4\tilde{x}\,a^{\alpha}(x)\Pi_{\alpha\beta}(x,\tilde{x})a^\beta(\tilde{x})+\ldots.
\end{split}\label{initial}
\end{equation}Here $f_{\alpha\beta} = \partial_\alpha a_\beta-\partial_\beta a_\alpha$ with $e^i=f^{0i}$ and  $b^i=\frac{1}{2}\epsilon^{ijk}f^{jk}$  referring to  the corresponding  electric and  magnetic fields.  In the action above $\Pi_{\alpha\beta}(x,\tilde{x})$ is the vacuum polarization tensor, whereas $+\ldots$ stands for  higher order terms in $a$ that can be safely ignored. Gauge and vacuum invariances can be used to determine the tensorial structure of the polarization tensor. In the presence of a constant  homogeneous magnetic field $\pmb{B}$ characterized by the tensor $\mathscr{F}_{\alpha\beta}=\partial_\alpha \mathscr{A}_\beta-\partial_\beta \mathscr{A}_\alpha$  with $\mathscr{F}_{0i}=0$ and $\mathfrak{F}=\frac{1}{2}B^2$, its covariant structure  reads \cite{Batalin}:
\begin{equation}\label{diagonaldecomposition}
\Pi_{\alpha\beta}(q)=\sum_{i=1}^3\varkappa_i\frac{\flat_{\alpha,i} \flat_{\beta,i}}{\flat_i^2},
\end{equation}Here $\varkappa_i$  with $i=1,2,3$ are the renormalized  $\Pi_{\alpha\beta}$-eigenvalues, which  are complex functions of the two invariants $q\tilde{\mathscr{F}}^2q/2\mathfrak{F}=q_\parallel^2-q_0^2$ and  $q\mathscr{F}^2q/2\mathfrak{F}=-q_\perp^2$.   Here $q_\perp$ ($q_\parallel$) is the  momentum perpendicular (parallel) to $\pmb{B}$, whereas $\tilde{\mathscr{F}}^{\alpha\beta}=\frac{1}{2}\epsilon^{\alpha\beta\mu\nu}\mathscr{F}_{\mu\nu}$ is the dual tensor of $\mathscr{F}_{\alpha\beta}$.  The associated  eigenvectors   $\flat_{\mu,1}=q^2\mathscr{F}_\mu^{\;\,\lambda}\mathscr{F}_{\lambda}^{\,\;\nu}q_\nu-(q\mathscr{F}^2q) q_\mu$, $\flat_{\mu,2}=\tilde{\mathscr{F}}_\mu^{\,\;\nu}q_\nu$ and $\flat_{\mu,3}=\mathscr{F}_\mu^{\,\;\nu}q_\nu$ are transverse to $q_\mu$ and  mutually orthogonal  [$\flat_{i}\flat_{j}=\delta_{ij}\flat_j^2$]. As a whole, they satisfy the completeness relation  $\mathpzc{g}_{\mu\nu}-q_\mu q_\nu/q^2=\sum_i \flat_{\mu,i}\flat_{\nu,i}/\flat_i^2$. These features allow us to express the effective Lagrangian of \eqref{initial} in a way that facilitates the establishment of the canonical  Hamiltonian $H=\int d^3x\,[\mathscr{H}-a_0\pmb{\nabla}\cdot{\pmb{\pi}}]$ with 
\begin{equation}\label{Hamiltonian}
\begin{split}
\mathscr{H}&=\int d^4\tilde{x}\left[\frac{1}{2}\pi_i(x)\varepsilon_{ij}^{-1}(x,\tilde{x})\pi_j(\tilde{x})\right.\\
&\qquad \qquad\qquad\qquad+\left.\frac{1}{2}b_i(x)\mu_{ij}^{-1}(x,\tilde{x})b_j(\tilde{x})\right].
\end{split}
\end{equation}The Hamiltonian resembles the one in dispersive and absorptive media \cite{Landau}. Indeed,  the canonical momentum  $\pmb{\pi}(x)=-\pmb{d}(x)$ is determined by the electric  displacement vector $d_i=\int d^4\tilde{x}\,\varepsilon_{ij}(x,\tilde{x})e_j(\tilde{x})$ with    $\varepsilon_{ij}(x,\tilde{x})=\int d^4q/(2\pi)^4\,\varepsilon_{ij}(q)\exp[-iq(x-\tilde{x})]$ denoting the dielectric tensor, the Fourier transform of which is
\begin{equation}\label{dielectrictensor}
\begin{split}
&\varepsilon_{ij}(q)=\left(1-\frac{\varkappa_1}{q^2}\right)\delta_{ij}+\frac{\varkappa_1-\varkappa_2}{q\tilde{\mathscr{F}}^2q}\frac{B_iB_j}{B^2}.
\end{split}
\end{equation}Likewise,  $\mu_{ij}^{-1}(x,\tilde{x})=\int d^4q/(2\pi)^4\,\mu_{ij}^{-1}(q)\exp[-iq(x-\tilde{x})]$ denotes the magnetic permeability tensor with 
\begin{equation}\label{susceptibilitytensor}
\begin{split}
&\mu_{ij}^{-1}(q)=\left(1-\frac{\varkappa_1}{q^2}\right)\delta_{ij}-\frac{\varkappa_1-\varkappa_3}{q\mathscr{F}^2q}\frac{B_iB_j}{B^2}.
\end{split}
\end{equation}  Notice that the complex nature of $\varkappa_i$ makes $\varepsilon_{ij}$ and $\mu_{ij}^{-1}$ non-Hermitian objects  in general, allowing for both dispersive and dissipative processes \cite{Shabad}. 

{\it Effective Thermal Approach}---Hereafter, we will use a thermodynamic equilibrium-based approach, which calls for real $\Pi_{\alpha\beta}$-eigenvalues and thus energy and momentum lying within the domain of  transparency \cite{ShabadJETP}, i.e., where the production of pairs does not occur. To establish the corresponding Helmholtz free energy $\Omega=-\beta^{-1}\ln\mathpzc{Z}$, the imaginary time formalism must be adopted, i.e.,  $t\to - i\tau$ with $0\leq \tau\leq \beta$ and $\beta=T^{-1}$. In this context,  $a_\mu(x)$ with $x_\mu=(\pmb{x},\tau)$ and $\mu=1,2,3,4$,  are promoted to  periodic functions in $\tau$ with period $\beta$. Likewise, the analytical continuation   $q_0\to iq_4$ with $q_4=2\pi n/\beta$ is carried out. Here, the partition function of our problem    $\mathpzc{Z}=\mathrm{Tr}\Big[\mathrm{T}\exp\big(-\int_0^\beta H_{\mathrm{ph}}(\tau)\big)\Big]$ is determined by the physical Hamiltonian $H_{\mathrm{ph}}=\int d^3x\,\mathscr{H}_{\mathrm{E}}$, which comprises the Euclidean version of Eq.~\eqref{Hamiltonian}.  We shall, however, start with the alternative representation 
\begin{equation}\label{principalZ}
\begin{split}
\mathpzc{Z}=&\mathrm{det}[\varepsilon_{ij}^{\mathrm{E}}(x,\tilde{x})]^{1/2}\oint \mathpzc{D}a_\mu\,\delta(\mathscr{G}[a_\mu])\\
&\qquad\qquad\times\mathrm{det}\left[\delta \mathscr{G}[^\chi a_\mu]/\delta\chi\right]_{\chi=0}\exp[-\Gamma_{\mathrm{E}}],
\end{split}
\end{equation}where $\Gamma_{\mathrm{E}}$ is the Euclidean variant of Eq.~\eqref{initial}, whereas  $\mathrm{det}\left[\delta \mathscr{G}[^{\chi}a_\mu]/\delta \chi\right]_{\chi=0}$ denotes the corresponding  Faddeev-Popov determinant.  Here,  $^\chi a_\mu=a_\mu+\partial_\mu \chi$ is the gauge transformed field. The weight factor  $\mathrm{det}[\varepsilon_{ij}^{\mathrm{E}}(x,\tilde{x})]^{1/2}$  arises as a result of integrating over $\pmb{\pi}$ in the phase-space formulation of the problem \cite{Bordag}. Calculations are simplified  when  $\mathscr{G}[a_\mu]=-\partial_\mu a_\mu/\sqrt{\xi}+\mathpzc{o}(x)$ with  an arbitrary function $\mathpzc{o}(x)$  of $x_\mu$ and  $\xi\in\mathbb{R}$.  Since  $\mathpzc{Z}$ does not depend  on the latter, it is  weighted   with   $\exp(-\frac{1}{2\xi}\int_0^\beta d\tau\int_V d^3x\, \mathpzc{o}^2)$, and  subsequently  integrated over  $\mathpzc{o}(x)$.   As a result,  the gauge-fixing choice  is  exponentiated,  and upon  integration over $a_\mu$ we end up with 
$\mathpzc{Z}=\prod_{q}[\beta^4(q_4^2+\omega_2^2(\pmb{q}))(q_4^2+\omega_3^2(\pmb{q}))]^{-\frac{1}{2}}$, where  $\omega_{2,3}^2(\pmb{q})=q_\parallel^2+f_{2,3}(q_\perp^2)$  are the general representations of  the  dispersion laws.    The obtained expression for $\mathpzc{Z}$ can be inserted into  the  free energy $\Omega$.  After summing  over  $q_4$,   and taking   $V\to\infty$, the free energy $\Omega=\Omega_{\mathrm{st}}+\Omega_{\mathrm{vac}}$ splits into  a  statistical contribution:
\begin{equation}
\begin{split}
\Omega_{\mathrm{st}}=\frac{V}{\beta}\sum_{i=2,3}\int \frac{d^3q}{(2\pi)^3}\ln\left(1-e^{-\beta\omega_{i}}\right),
\end{split}\label{StatFreeEnergy}
\end{equation} and a  vacuum term  $\Omega_{\mathrm{vac}}=\frac{1}{2}V\sum_{i}\int \frac{d^3q}{(2\pi)^3}\omega_{i}$, which is divergent. The latter can be regularized  using  the proper time method  so that   $\Omega_{\mathrm{vac}}\to-\frac{1}{32\pi^2}V\sum_{i=2,3}\int_0^{\infty} d q_\perp^2 \int_{1/\Lambda^2}^{\infty}\frac{d\tau}{\tau^2}[e^{-f_i(q_\perp^2)\tau}-e^{-q_\perp^2\tau}]$ with  $\Lambda$ denoting the cutoff  parameter.  We remark that these formulae are independent of any approximation required to determine the polarization tensor and that the associated  equations of state  $\mathpzc{P}_\parallel V=-\Omega$ and $\mathpzc{P}_\perp V=-\Omega-\mathpzc{M} B$---with $\mathpzc{M}(B)=-\partial\Omega/\partial B$  the magnetization of  the photon ensemble \cite{mmph,mmps,Villalba-Chavez:2012pmx}---manifest the  anisotropy induced by the external magnetic field $\pmb{B}$. 

{\it Anisotropic Blackbody Radiation}---From now on we will focus on the phenomenological repercussions of Eq.~\eqref{StatFreeEnergy}  by using the  one-loop expression of  $\Pi_{\alpha\beta}$ \cite{Batalin,Baier,Wu,Urrutia,Hattori1,Hattori2,Felix}.   The  $\varkappa_i$ can generally be  expressed as sums over the Landau levels ($n,n^\prime$) that the  mutually independent  fermions in the loop can occupy.  We shall focus, however,  on  the asymptotic region of strong  magnetic fields $\mathfrak{b}\gg1$ with $\mathfrak{b}=B/B_{\mathrm{cr}}$,  and  $m^2 \mathfrak{b}\gg q_0^2-q_\parallel^2$. In this case, contributions from doubly excited levels  ($n,n^\prime \geqslant 1$) are suppressed, and  the leading order contributions of $\varkappa_{1}$ and $\varkappa_{3}$ for $q_\perp\ll m\sqrt{2\mathfrak{b}}$  arise from cases in which  the loop includes one fermion in  an excited state ($n=0,n^\prime=1$ or $n=1,n^\prime=0$ ) \cite{ShabadJETP,Ferrer}:
\begin{equation}\label{eigenvalues13}
\begin{split}
&\varkappa_1(q_0^2-q_\parallel^2,q_\perp)=\frac{\alpha}{3\pi}q^2\left[\ln\left(\frac{\mathfrak{b}}{\gamma\pi}\right)-0.065\right],\\
&\varkappa_3(q_0^2-q_\parallel^2,q_\perp)=\varkappa_1+\frac{\alpha}{3\pi}q_\perp^2,
\end{split}
\end{equation}  In contrast,  $\varkappa_{2}$ is determined by the situation in which both leptons occupy  the lowest Landau level $n=n^\prime=0$:
\begin{equation}\label{eigenvalue2}
\begin{split}
&\varkappa_2(q_0^2-q_\parallel^2,q_\perp)=\frac{2\alpha m^2 \mathfrak{b}}{\pi}\exp\left(-\frac{q_\perp^2}{2 m^2 \mathfrak{b}}\right)\\
&\qquad\qquad \times \left[\frac{4m^2\arctan\left(\sqrt{\frac{q_0^2-q_\parallel^2}{4m^2+q_\parallel^2-q_0^2}}\right)}{\sqrt{(q_0^2-q_\parallel^2)(4m^2+q_\parallel^2-q_0^2)}}-1\right].
\end{split}
\end{equation} Here $\ln(\gamma)=0.577\ldots$ is the Euler constant and   $\alpha=e^2/(4\pi)$ is  the fine structure constant. Observe that $\varkappa_2$, in contrast to $\varkappa_{1,3}$,  is   singular  at the first pair creation threshold $q_0^2-q_\parallel^2=4m^2$, which causes a strong  birefringence near the threshold. 

\begin{figure}\hspace{.3cm}
\includegraphics[width=0.208\textwidth]{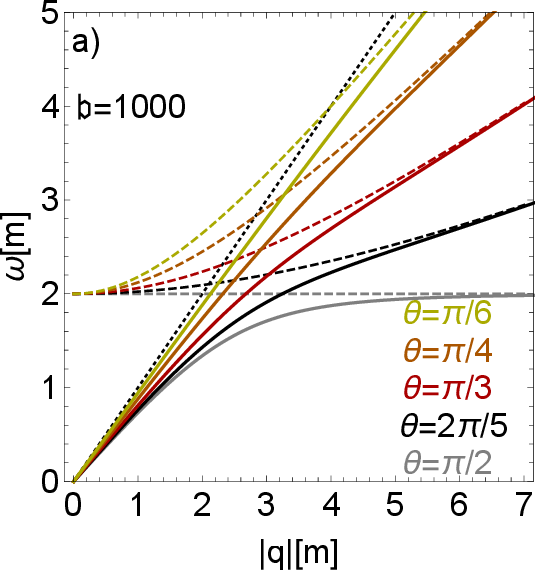}\hspace{0.3cm}
\includegraphics[width=0.228\textwidth]{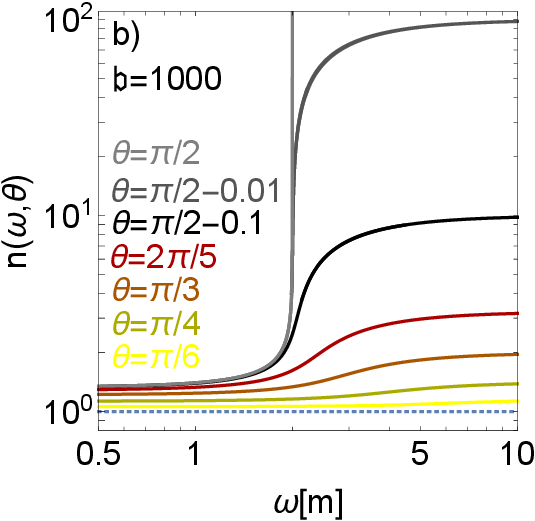}
\includegraphics[width=0.235\textwidth]{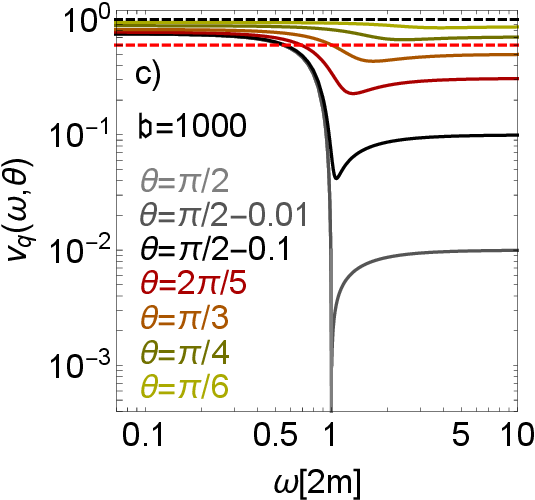}
\includegraphics[width=0.237\textwidth]{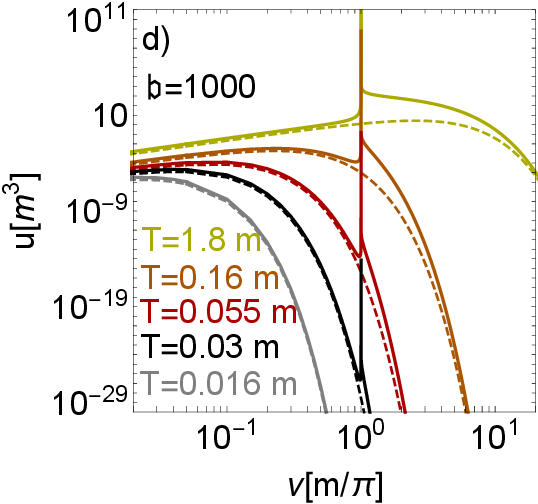}
\caption{(a)  Dispersion relations. The diagonal dotted line is linked to  mode 3, whereas the solid curves belong to  mode 2. The dashed curves follow from  the first pair creation threshold $\omega=(\pmb{q}^2\cos^2(\theta)+4m^2)^{1/2}$.  Curves sharing a color are linked to a common $\theta\in[0,\pi]$. (b) Dependence of the refraction indices $\mathpzc{n}_i(\omega,\theta)$ on $\omega$ for various angles $\theta$. (c) Behavior of the group velocity $\mathpzc{v}_{\pmb{q},i}(\omega,\theta)$ with $\omega$ for various angles $\theta$. The horizontal red dashed line shows the velocity needed to escape from a NS  with $R_\star\approx 10\;\rm Km$, $M_\star\approx 1.4 M_{\odot}$. (d) Blackbody spectrum of the second  (solid) and third (dashed) modes. Curves sharing a color are linked to a common temperature.} 
\label{fig:2}
\end{figure}

Hereafter, we will restrict ourselves to  field strengths $1\ll \mathfrak{b}<3\pi/\alpha$. Then  $\omega_3(\pmb{q})\approx \vert \pmb{q}\vert$,  whereas  the solution of $q^2=\varkappa_2$ must be found numerically.  Fig.~\ref{fig:2}(a) shows the  behavior of $\omega_2(\pmb{q})$ for different values of the angle  $\theta=\measuredangle(\pmb{q},\pmb{B})$.  A key consequence of the singularity of $\varkappa_2$  is the significant deviation of the dispersion curve---in solid style---from the light cone law---diagonal dotted line---which occurs despite  the distinctive smallness of  $\alpha$ \cite{ShabadAstroSpaSci}.  The described behavior has a direct impact on the quantum vacuum's refraction properties.   While mode-3 photons have a refractive index close to the classical value $\mathpzc{n}_3=\vert \pmb{q}\vert/\omega_3\approx 1$,  mode-2 quanta are characterized by  $\mathpzc{n}_2(\omega,\theta)=\vert \pmb{q}\vert/\omega_2$ that tends to grow unlimited as $\theta\to \pi/2$  [see Fig.~\ref{fig:2}(b)].  The mode-2 dispersion law also  influences  the group velocity $\pmb{\mathpzc{v}}_2=\pmb{\nabla}_{\pmb{q}}\omega_2$, which differs from the phase velocity $\pmb{v}_{\mathrm{ph},2}=\frac{\omega_2}{\vert\pmb{q}\vert}\pmb{n}$ with $\pmb{n}=\pmb{q}/\vert\pmb{q}\vert$ in both  magnitude and  direction. Notably, the component of $\pmb{\mathpzc{v}}_2$ along the $\pmb{q}$-direction, i.e., $\pmb{\mathpzc{v}}_{\pmb{q},2}$  reaches values smaller than the speed of light in vacuum [$\vert\pmb{\mathpzc{v}}_{\pmb{q},2}\vert<1$] when $\omega>2m$ [solid curves in  Fig.~\ref{fig:2}(c)].  We note that the speed's falling of mode-2 becomes  significant as $\theta\to\pi/2$. This  contrasts with  $\vert\pmb{\mathpzc{v}}_3\vert\approx1$ [black dashed line].
  
\begin{figure}
\includegraphics[width=0.237\textwidth]{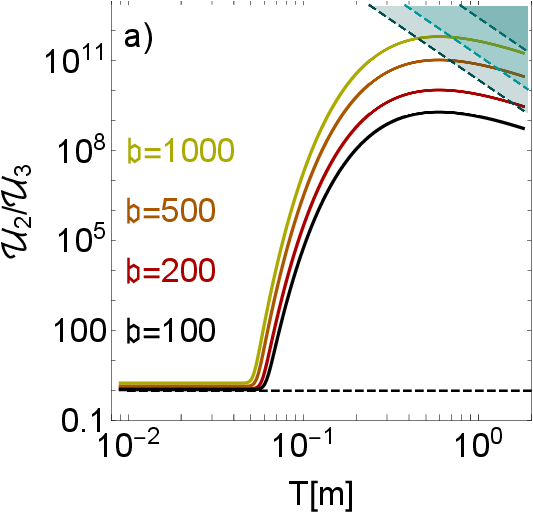}
\includegraphics[width=0.235\textwidth]{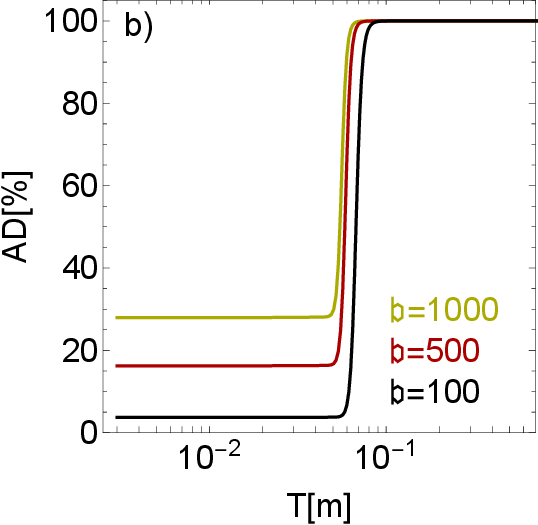}
\includegraphics[width=0.237\textwidth]{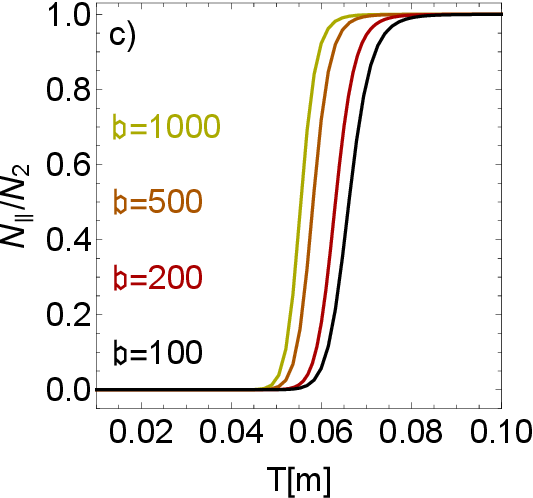}
 \includegraphics[width=0.237\textwidth]{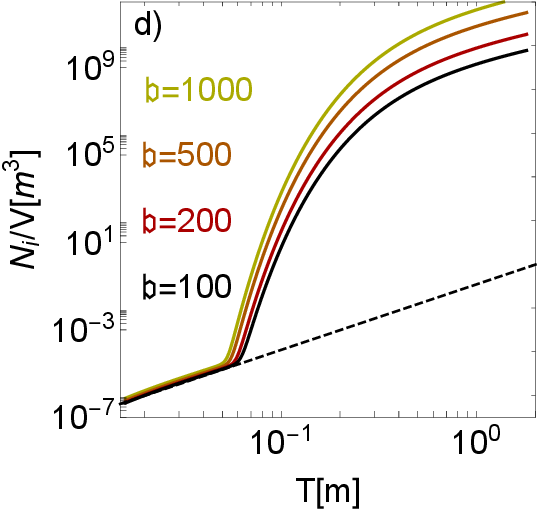}
\caption{(a) Temperature dependence of the ratio $\mathpzc{U}_2/\mathpzc{U}_3$ between the internal energy density due to the second and third propagating modes  for various field strengths.  The  black dashed line gives for comparison a ratio of unity.  The shaded sectors show where a condensate of photons either destabilizes a NS or leads to a violation of causality.   (b) Asymmetry  degree of heat radiation in a magnetized vacuum vs temperature for different $\mathfrak{b}$. (c) Condensate fraction of mode-2 photons occupying slow light states as a function of temperature for various fields. (d) Phase diagram for heat radiation. The dashed line describes the mean number of mode-3 photons.}
\label{fig:3}
\end{figure}

We aim to evaluate how the described birefringence phenomenon affects Planck's radiation law.  To  this end, we primarily investigate the internal energy density $\mathpzc{U}=\frac{1}{V}\frac{\partial(\beta \Omega)}{\partial\beta}=\sum_i\mathpzc{U}_i$. Going over to spherical variables  
\begin{equation}\label{MPRL}
\begin{split}
u_i&=\frac{d\mathpzc{U}_i}{d\nu}=\frac{4\pi^2\nu^3}{e^{2\pi\beta\nu}-1}\int_0^\pi d\theta \frac{\mathpzc{n}_{i}^2\sin\theta}{\vert\pmb{\mathpzc{v}}_{\pmb{q},i}\vert}.
\end{split}
\end{equation}  Here, the  factor  $1/\vert\pmb{\mathpzc{v}}_{\pmb{q},i}\vert$  is the Jacobian resulting from  adopting  $\nu=\omega/2\pi$  as integration variable. Since $\mathpzc{v}_{\pmb{q},2}=0$ at $\theta=\pi/2$, this operation renders the corresponding  density of states  (DOS) singular, mirroring  the van Hove singularities  in solids \cite{vanHove}. Calculations were then carried out by introducing a physical cutoff  for  $q_\perp^2$, which is naturally  provided  by the factor $2 m^2\mathfrak{b}$ present in the exponent of $\varkappa_2$ \cite{ShabadAstroSpaSci,Ferrer}.  The shown frequency range $\nu_{\mathrm{L}}<\nu<\nu_{\mathrm{H}}$  covers photon frequencies above $\nu_{\mathrm{L}}=10^{-2} m/\pi$ but below  the lowest value of  the second pair production threshold $\nu_{\mathrm{H}}\approx\sqrt{2\mathfrak{b}} m/2\pi$.  We remark that,  for $B\to\infty$, this threshold is moved to remote energy values, which favors the equilibrium approach. Aside from these details, Eq.~\eqref{MPRL} shares certain similarities with Planck's radiation law for dispersive anisotropic media  \cite{Cole,Mercier,Eric}.   Fig.~\ref{fig:2}(d) summarizes the  behavior of  $u_{i}$ for   $\mathfrak{b}=1000$.   As one could anticipate, the  spectrum of the second mode (solid curves) shows a remarkable departure  at  $\nu=m/\pi$,  exhibiting a narrow resonant peak in which the quantum degeneracy enhances with increasing temperature.  Indeed, the study suggests that a crossover occurs around a certain temperature $0.04\, m<T_*< 0.06\, m$ above which the peak of the latter exceeds Wien's maximum. The existence of  $T_*\ll 2m$ is confirmed by the outcomes presented in Fig.~\ref{fig:3}, which we discuss below. We emphasize that  the origin of the resonance is rooted in the deviation exhibited by the dispersion relation which can be approximated by  $\omega_2\approx (q_\parallel^2+4m^2)^{1/2}\approx q_\parallel^2/4m+2m$  at $\theta\sim \pi/2$  [see Fig.~\ref{fig:2}(a)]. Notably,  $\gamma$-quanta populating this resonance---characterized by $q_\parallel\ll 2m$ and $q_\perp \gg 2m$---have group velocities $\vert\pmb{\mathpzc{v}}\vert\approx\mathpzc{v}_\parallel\ll 1$, rendering the thermal wave packet a slow  light state with a low-dimensional energy transport along $\pmb{B}$.

Observe that the results  in Fig.~\ref{fig:2}(d) have been obtained for $T<2m \ll \sqrt{\vert eB\vert}$.  Notably,  the right tail  of the resonance also exceeds  that of mode-3 photons. This feature indicates that the contribution of mode-2 photons  to $\mathpzc{U}$  outweighs the one due to the third mode.  Figure~\ref{fig:3}(a) exhibits the temperature dependence of $\mathpzc{U}_2/\mathpzc{U}_3$ which determines  $\mathpzc{U}$ in units of  $\mathpzc{U}_3\approx \pi^2 T^4/30$. For $T<T_{*}$, $\mathpzc{U}\approx(1+\varepsilon_\parallel/\varepsilon_\perp)\mathpzc{U}_3$,  where---for $1\ll\mathfrak{b}<3\pi/\alpha\approx 1.3\times 10^3$---$\varepsilon_\perp\approx 1$ and $\varepsilon_\parallel\approx1-\alpha\mathfrak{b}/3\pi$ are the infrared  approximations of  $\varepsilon_{ij}$ eigenvalues \cite{ShabadLangrangian,Villalba-Chavez:2012pmx}. The $T$ scaling of $\mathpzc{U}_{2}$ for  $T>T_{*}$  is  however  significantly stronger than $\mathpzc{U}_3\sim  T^4$, outweighing the contribution of the gas made by mode-3 photons  by  various   orders of magnitude.   This  indicates that vastly more mode-2 than mode-3 photons exist.  Figure~\ref{fig:3}(b) confirms this outcome via  the asymmetry degree  $AD=(N_{2}-N_3)/(N_2+N_3)$ with  $N_{i}=2\pi V\int_{\nu_{\mathrm{L}}}^{\nu_{\mathrm{H}}}\frac{d\nu\; \nu^2}{\exp(2\pi\nu/T)-1}\int_0^\pi \frac{d\theta\,\sin(\theta)}{\vert\mathpzc{v}_{\pmb{q},i}\vert}\mathpzc{n}_i^2$   the mean  number of  mode-$i$ photons.. For $T>T_{*}$  an almost  complete asymmetry   $\approx100\%$  is reached between the two photon species regardless of $B$. At low temperatures $T<T_*$, the results indicate that  $N_2$ exceed $N_3$ by increasing $B$. 

The fraction  of slow $\gamma$ quanta  $N_{\parallel}/N_2$  that occupies the resonance---$\mathpzc{v}_\parallel\ll1,\,\mathpzc{v}_\perp\approx0$---is depicted  in Fig.~\ref{fig:3}(c). Here, $N_{\parallel}$ was determined by choosing the lowest integration limit as  the spectral point at $T\approx T_*$ (see  the red curve in Fig.~\ref{fig:2}(d)) where the slope transitions from negative to positive. In contrast,  the upper integration limit was defined at the point on the right resonance tail that is at the same height as the lower limit.   The  established integration limits were then used to determine the density of particles at other temperatures. Figure~\ref{fig:3}(c) shows that  for $T > T_{*}$,  the population of  this thermal state occurs copiously, resembling  a many-body condensate.  We remark, however, that this is not a  BEC because  the  divergence in the spectrum is  driven by a  kinematic factor.  Indeed, the described pileup of thermal  photons is attributed to the combined effects of temperature and the aforementioned  singularity.  The crossover and condensation of the slow thermal light state are also illustrated in Fig.~\ref{fig:3}(d), which shows how the photon number evolves with  $T$.  On the left side of  $T_*$, $N_2$ shows a $T$-dependence, similar to  $N_3 \propto T^3$. However, on the right of $T_*$, the dependence on $T$ is no longer of the form $T^3$.  As a result, at $T\approx 0.18\, m\sim 10^{9}\, \rm K$ and $\mathfrak{b}=1000$, the density $N_2/V$ exceeds  $N_3/V\sim10^{28} \,\rm cm^{-3}$  by ten orders of magnitude.  

\emph{Astrophysical implications}---The results suggest that NS cores with $10^9\, \mathrm{K}\lesssim T\lesssim 10^{11}\,\rm K$ and $10^{14}\,\mathrm{G}\lesssim B\lesssim10^{16}\,\rm G$ can sustain the predicted condensate of $\gamma$ photons. Even if $\pmb{B}$ is inhomogeneous in such a context, its direction and magnitude could remain uniform over coherence domains larger than $\lambda^3$ with $\lambda=(2m)^{-1}\sim 0.1\, \rm pm$. Partitioning the space into these domains enables us to infer qualitative characteristics of the condensate that stem from the curved $\pmb{B}$-geometry. An immediate consequence of this sort of ``locally constant field approximation'' is that the slow $\gamma$ photons gather and move along the curved lines of force. This process emulates the photon-capture effect predicted for curvature radiation in pulsar magnetospheres, a phenomenon central to their emission mechanism \cite{ShabadNature,ShabadAstroSpaSci,Herold,Bathia,Selym2}. However, unlike curvature radiation, the thermal quanta in the condensate experience a high $\mathpzc{n}_2$, making their escape from the strong $B$-regions improbable.  These conditions promote total internal reflection of $\gamma$ radiation when it  attempts to traverse from a strong (high) to a weak (low) field ($\mathpzc{n}_2$) domain. This retention process is even more favorable, as the captured photons move with a velocity  $\mathpzc{v}_{\mathrm{rms}}=\sqrt{\langle\mathpzc{v}_2^2\rangle}\sim 10^{-10}$ insufficient to overcome the star's escape velocity  $v_{\mathrm{esc}}=\sqrt{2G M_\star/R_\star}\approx 0.5$   ($M_\star=1.4 M_{\odot}$ and $R_\star=10\,\rm km$) [red  dashed line in  Fig.~\ref{fig:2}(c)].  As a result,  the condensate freezes in the  medium  while mode-3 photons propagate along straight lines and decay---high environment opacity---via  $\gamma_3\to\gamma_2^\prime+\gamma_2^{\prime\prime}$ \cite{ShabadAstroSpaSci,Holger,AdlerPRL1}.

Now, the inner composition of NSs remains a challenging open problem. Depending on the core density, different phases may occur. Near nuclear saturation, $\varrho_{\mathrm{s}}=2.8\times 10^{14}\,\rm g/cm^3$  [$N_{\mathrm{s}}/V=1.7\times 10^{38}\, \rm cm^{-3}$]  the matter is almost entirely neutrons, with only a small proton fraction electrically  neutralized by electrons \cite{Prakash,Chamel,shapiro}. We note that,  at $T \approx 3.6\times 10^{9}\,\rm K$ and $B \approx 2.2 \times 0^{16}\,\rm G$,  $N_2/N_{\mathrm{s}}\sim 10^3$.  Also,  the internal energy in the condensate  $\mathpzc{U}_2\approx 6.4\times  10^{34}\, \rm erg/cm^3$ is comparable to  $\mathpzc{U}_{\mathrm{nuc}}\sim \varrho_{\mathrm{s}} =2.5 \times 10^{35}\,\rm erg/cm^{3}$.  This implies  that the star's mass  $M_\star\approx \frac{4}{3}\pi R_\star^3\mathpzc{U}_{\mathrm{nuc}}[1+(\mathpzc{U}_{2}+\mathpzc{U}_{3})/\mathpzc{U}_{\mathrm{nuc}}]$ is not only determined by the nuclear matter but also by  the condensate.  The mass $M_\star$ should however remain below the stability limit, $M_\star\leqslant 2.3\,M_\odot$; otherwise, gravitational collapse is triggered \cite{Prakash}.  Thus, a condensate of $\gamma$ photons in a NS core keeps the star stable as long as $\mathpzc{U}_2/\mathpzc{U}_3\leqslant(\mathpzc{U}_{\mathrm{max}}- \mathpzc{U}_{\mathrm{nuc}})/\mathpzc{U}_3-1$ with $\mathpzc{U}_{\mathrm{max}}=6.9M_{\odot}/(4\pi R_\star^3)$. Taking  $R_\star=10\,\rm km$, $\mathpzc{U}_{\mathrm{max}}\approx9\times 10^{35}\,\rm erg/cm^3$ results---we delineated the region in Fig.~\ref{fig:3}(a) where the condensate renders a NS unstable [intermediate shaded sector].

%However, massive NS cores could reach even higher nuclear densities leading to  $\mathpzc{U}_{\mathrm{nuc}}\sim 10^{37}\,\rm erg/cm^3$ .

A highest  $\mathpzc{U}_{\mathrm{max}}\approx 1.2\times 10^{37}\,\rm erg/cm^3$---independent of $R_\star$---has been established for $M_\star=2.2 M_\odot$  by requiring  the sound speed to not exceed the speed of light in vacuum (causality) \cite{PRLPrakash}.  The unphysical region resulting from this constraint turns out to be smaller, maintaining  $\mathpzc{U}_{\mathrm{nuc}}$ as in the previous case [smaller shaded sector]. However, this region expands as $\mathpzc{U}_{\mathrm{nuc}}\to \mathpzc{U}_{\mathrm{max}}$: for  $\mathpzc{U}_{\mathrm{nuc}}=1.19\times 10^{37}\,\rm erg/cm^3$---in line with the  density of massive NS cores \cite{Haensel}---the  shaded sector in  Fig.~\ref{fig:3}(a) becomes larger.  Notably,  the intersections of the intermediate  cyan dashed line with the solid curves indicate that, the stronger  $B$ is, a stable NS core can only exist for lower  $T$-values.   We note that, although the condensate deepens the gravitational potential, its contribution to the pressure supporting hydrostatic equilibrium remains negligible compared to the influence of nuclear matter, i.e., $\mathpzc{P}_{\mathrm{nuc}}\gg\mathpzc{P}_{\parallel,\perp}$ with $\mathpzc{P}_{\mathrm{nuc}}\sim(10^{35}-10^{36})\,\rm erg/cm^3$. Moreover, its presence in the NS core would not accelerate stellar cooling via  neutrino-antineutrino pair production, since this process---like the creation of electron-positron pairs---is kinematically forbidden for the photons in the condensate.  However, with the star's cooling, the phase is expected to dilute and the photon ensemble eventually crosses over into the standard blackbody phase before its thermal extinction.

 {\it Conclusions}---The cyclotron resonance near the first pair-creation threshold induces a van Hove singularity in the DOS of mode 2 photons.  When the system's temperature  $T>T_*$, the singularity impedes thermal redistribution and promotes macroscopic occupation of the singular state. The described mechanism is universal, arising from Bose-Einstein statistics combined with a divergent DOS and  renders the  photon ensemble to behave like a magnetized, compressible, slow boson fluid.  We have revealed  that the condensate of $\gamma$ photons  could exist in the interior of NSs making  the star more compact. Temperatures and fields for which its presence would induce instability   or violate causality have been ruled out. \\

%%%%%%%%%%%%%%%%%%%%%%%%%%%%%%%%%%%%%%%%%%%%%%%%%%%%%%%%%%%
{\it Acknowledgments}--SVC thanks Reinhard Alkofer for an interesting discussion on path integral quantization in nonlocal theories. RE acknowledges funding by the Deutsche Forschungsgemeinschaft (DFG, German Research Foundation), under Projektnummer 277101999 -- TRR 183 (project C01),  and under Germany's Excellence Strategy -- Cluster of Excellence Matter and Light for  Quantum Computing (ML4Q) EXC 2004/1 -- 390534769.  CM thanks DFG for funding of Grant No. 392856280 within the Research Unit FOR 2783/2 ``Probing the Quantum Vacuum at the High-Intensity Frontier''.


\begin{thebibliography}{93}


\expandafter\ifx\csname
natexlab\endcsname\relax\def\natexlab#1{#1}\fi
\expandafter\ifx\csname bibnamefont\endcsname\relax
  \def\bibnamefont#1{#1}\fi
\expandafter\ifx\csname bibfnamefont\endcsname\relax
  \def\bibfnamefont#1{#1}\fi
\expandafter\ifx\csname citenamefont\endcsname\relax
  \def\citenamefont#1{#1}\fi
\expandafter\ifx\csname url\endcsname\relax
  \def\url#1{\texttt{#1}}\fi
\expandafter\ifx\csname urlprefix\endcsname\relax\def\urlprefix{URL
}\fi \providecommand{\bibinfo}[2]{#2}
\providecommand{\eprint}[2][]{\url{#2}}

%%%%%%%%%%%%%%%%%%%%%%%%INTRODUCTION%%%%%%%%%%%%%%%%%%%%%%%%%%

\bibitem{KlaersNatPhys} 
\bibinfo{author}{\bibfnamefont{J.~}\bibnamefont{Klaers}}, \bibinfo{author}{\bibfnamefont{F.~}\bibnamefont{Vewinger}}, and \bibinfo{author}{\bibfnamefont{M.~}\bibnamefont{Weitz}}, 
\bibinfo{title}{Thermalization of a two-dimensional photonic gas in a 'white wall' photon box},
\bibinfo{journal}{Nat. Phys.} \pmb{\bibinfo{volume}{6}}, 
\bibinfo{pages}{512}  (\bibinfo{year}{2010}).

\bibitem{Klaers} 
\bibinfo{author}{\bibfnamefont{J.~}\bibnamefont{Klaers}}, \bibinfo{author}{\bibfnamefont{J.~}\bibnamefont{Schmitt}}, \bibinfo{author}{\bibfnamefont{F.~}\bibnamefont{Vewinger}}, and \bibinfo{author}{\bibfnamefont{M.~}\bibnamefont{Weitz}}, 
\bibinfo{title}{Bose-Einstein condensation of photons in an optical microcavity},
\bibinfo{journal}{Nature (London)} \pmb{\bibinfo{volume}{468}}, 
\bibinfo{pages}{545}  (\bibinfo{year}{2010}).

\bibitem{KlaersNatComm} 
\bibinfo{author}{\bibfnamefont{T.~}\bibnamefont{Damm}}, \bibinfo{author}{\bibfnamefont{J.~}\bibnamefont{Schmitt}},  \bibinfo{author}{\bibfnamefont{Q.~}\bibnamefont{Liang}},   \bibnamefont{\emph{et al.}}
\bibinfo{title}{Calorimetry of a Bose-Einstein-condensed photon gas},
\bibinfo{journal}{Nat. Commun.} \pmb{\bibinfo{volume}{7}}, 
\bibinfo{pages}{11340}  (\bibinfo{year}{2016}).

\bibitem{Weill1} 
\bibinfo{author}{\bibfnamefont{R.~}\bibnamefont{Weill}}, \bibinfo{author}{\bibfnamefont{A.~}\bibnamefont{Bekker}},  \bibinfo{author}{\bibfnamefont{B.~}\bibnamefont{Levit}},  \bibnamefont{and}
\bibinfo{author}{\bibfnamefont{B.}~\bibnamefont{Fischer}}, 
\bibinfo{title}{Bose-Einstein condensation of photons in an erbium-ytterbium co-doped fiber cavity},
\bibinfo{journal}{Nat. Commun.} \pmb{\bibinfo{volume}{10}}, 
\bibinfo{pages}{747}  (\bibinfo{year}{2019}).

\bibitem{Weill2} 
\bibinfo{author}{\bibfnamefont{R.~}\bibnamefont{Weill}}, \bibinfo{author}{\bibfnamefont{A.~}\bibnamefont{Bekker}},  \bibinfo{author}{\bibfnamefont{B.~}\bibnamefont{Levit}},  \bibnamefont{and}
\bibinfo{author}{\bibfnamefont{B.}~\bibnamefont{Fischer}}, 
\bibinfo{title}{Bose-Einstein condensation of photons in a long fiber cavity},
\bibinfo{journal}{Opt. Express} \pmb{\bibinfo{volume}{29}}, 
\bibinfo{pages}{27807}  (\bibinfo{year}{2021}).

\bibitem{Barland} 
\bibinfo{author}{\bibfnamefont{S.~}\bibnamefont{Barland}}, \bibinfo{author}{\bibfnamefont{P.~}\bibnamefont{Azam}}, \bibinfo{author}{\bibfnamefont{G.~L.~}\bibnamefont{Lippi}},  \bibnamefont{\emph{et al.}}
\bibinfo{title}{Photon thermalization and a condensation phase transition in an electrically pumped semiconductor microresonator}, \bibinfo{journal}{Opt. Express} \pmb{\bibinfo{volume}{29}}, 
\bibinfo{pages}{8368}  (\bibinfo{year}{2021}).

\bibitem{Schofield} 
\bibinfo{author}{\bibfnamefont{R.~C.}~\bibnamefont{Schofield}}, \bibinfo{author}{\bibfnamefont{M.~}\bibnamefont{Fu}}, \bibinfo{author}{\bibfnamefont{E.~}\bibnamefont{Clarke}},  \bibnamefont{\emph{et al.}} 
\bibinfo{title}{Bose-Einstein condensation of light in a semiconductor quantum well microcavity}, 
\bibinfo{journal}{Nat. Commun.} \pmb{\bibinfo{volume}{18}}, 
\bibinfo{pages}{1083}  (\bibinfo{year}{2024}).

\bibitem{Batalin}
\bibinfo{author}{\bibfnamefont{I.~A.}~\bibnamefont{Batalin}}  
\bibnamefont{and}
\bibinfo{author}{\bibfnamefont{A.~E.}~\bibnamefont{Shabad}}, 
\bibinfo{Title}{Green's function of a photon in a constant homogeneous electromagnetic field of general form},
\bibinfo{journal}{Sov. Phys. JETP} \pmb{\bibinfo{volume}{60}}, 483 (1971);  [\bibinfo{journal}{Zh. \'Eksp. Teor. Fiz.} \pmb{\bibinfo{volume}{60}}, \bibinfo{pages}{894} (\bibinfo{year}{1971})].

\bibitem{Wu}
\bibinfo{author}{\bibfnamefont{Wu~Yang}~\bibnamefont{Tsai}}, 
\bibinfo{Title}{Vacuum polarization in homogeneous magnetic fields},
\bibinfo{journal}{Phys. Rev. D} \pmb{\bibinfo{volume}{10}},
\bibinfo{pages}{2699}  (\bibinfo{year}{1974}).

\bibitem{Baier}
\bibinfo{author}{\bibfnamefont{V.~N.}~\bibnamefont{Baier}}, 
\bibinfo{author}{\bibfnamefont{V.~M.}~\bibnamefont{Katkov}}, 
\bibnamefont{and}
\bibinfo{author}{\bibfnamefont{V.~M.}~\bibnamefont{Strakhovenko}}, 
\bibinfo{Title}{Operator approach to quantum electrodynamics in an external field. Electron loops},
\bibinfo{journal}{Sov. Phys. JETP}  \pmb{\bibinfo{volume}{41}},
\bibinfo{pages}{198}  (\bibinfo{year}{1975}); [\bibinfo{journal}{Zh. \'Eksp. Teor. Fiz.} \pmb{\bibinfo{volume}{68}},
\bibinfo{pages}{198} (\bibinfo{year}{1975})].

\bibitem{Urrutia}
\bibinfo{author}{\bibfnamefont{L.~F.}~\bibnamefont{Urrutia}}, 
\bibinfo{Title}{Vacuum polarization in parallel homogeneous electric and magnetic fields},
\bibinfo{journal}{Phys. Rev. D} \pmb{\bibinfo{volume}{17}},
\bibinfo{pages}{1977}  (\bibinfo{year}{1978}).

\bibitem{Hattori1}
\bibinfo{author}{\bibfnamefont{K.}~\bibnamefont{Hattori}}~\bibnamefont{and}~\bibinfo{author}{\bibfnamefont{K.}\bibnamefont{Itakura}}, 
\bibinfo{Title}{Vacuum birefringence in strong magnetic fields: (I) Photon polarization tensor with all the Landau levels},
\bibinfo{journal}{Ann. Phys. (Amsterdam)} \pmb{\bibinfo{volume}{330}},
\bibinfo{pages}{23} (\bibinfo{year}{2013}).

\bibitem{Hattori2}
\bibinfo{author}{\bibfnamefont{K.}~\bibnamefont{Hattori}}~\bibnamefont{and}~\bibinfo{author}{\bibfnamefont{K.}\bibnamefont{Itakura}}, 
\bibinfo{Title}{Vacuum birefringence in strong magnetic fields: (II) Complex refractive index from the lowest Landau level},
\bibinfo{journal}{Ann. Phys. (Amsterdam)} \pmb{\bibinfo{volume}{334}},
\bibinfo{pages}{58} (\bibinfo{year}{2013}).

\bibitem{Felix}
\bibinfo{author}{\bibfnamefont{F.}~\bibnamefont{Karbstein}}, 
\bibinfo{Title}{The photon polarization tensor in a homogeneous magnetic or electric field},
\bibinfo{journal}{Phys. Rev. D} \pmb{\bibinfo{volume}{88}},
\bibinfo{pages}{085033} (\bibinfo{year}{2013}).

\bibitem{ShabadLett}
\bibinfo{author}{\bibfnamefont{A.~E.}~\bibnamefont{Shabad}}, 
\bibinfo{Title}{Cyclotronic resonance in the vacuum polarization},
\bibinfo{journal}{Lett. Nuovo Cimento} \pmb{\bibinfo{volume}{2}},
\bibinfo{pages}{457} (\bibinfo{year}{1972}).

\bibitem{Shabad}
\bibinfo{author}{\bibfnamefont{A.~E.}~\bibnamefont{Shabad}}, 
\bibinfo{Title}{Photon dispersion in a strong magnetic field},
\bibinfo{journal}{Ann. Phys.} \pmb{\bibinfo{volume}{90}},
\bibinfo{pages}{166} (\bibinfo{year}{1975}).

\bibitem{Witte}
\bibinfo{author}{\bibfnamefont{N.~S.}~\bibnamefont{Witte}}, 
\bibinfo{Title}{Polarization of the magnetized scalar and spinor vacua},
\bibinfo{journal}{J. Phys. A} \pmb{\bibinfo{volume}{23}},
\bibinfo{pages}{5257} (\bibinfo{year}{1990}).

\bibitem{ShabadJETP}
\bibinfo{author}{\bibfnamefont{A.~E.}~\bibnamefont{Shabad}}, 
\bibinfo{Title}{Photon propagation in a supercritical magnetic field},
\bibinfo{journal}{Sov. Phys. JETP} \pmb{\bibinfo{volume}{98}},
\bibinfo{pages}{186} (\bibinfo{year}{2004}).

\bibitem{Ashcroft}
\bibinfo{author}{\bibfnamefont{N.}~\bibnamefont{Ashcroft}} \bibnamefont{and}
\bibinfo{author}{\bibfnamefont{N.}~\bibnamefont{Mermin}}, 
\bibinfo{Title}{Solid State Physics},
\bibinfo{journal}{Harcourt Inc.}, (\bibinfo{year}{1976}).

\bibitem{AdlerPRL1}
\bibinfo{author}{\bibfnamefont{S.~L.}~\bibnamefont{Adler}}, 
\bibinfo{author}{\bibfnamefont{J.~N.}~\bibnamefont{Bahcall}}, 
\bibinfo{author}{\bibfnamefont{G.~G.}~\bibnamefont{Callan}}, 
\bibnamefont{and}
\bibinfo{author}{\bibfnamefont{M.~N.}~\bibnamefont{Rosenbluth}}, 
\bibinfo{Title}{Photon splitting in a strong magnetic field},
\bibinfo{journal}{Phys. Rev. Lett.} \pmb{\bibinfo{volume}{25}},
\bibinfo{pages}{1061} (\bibinfo{year}{1970}).

\bibitem{Adler:1971wn}
\bibinfo{author}{\bibfnamefont{S.~L.}~\bibnamefont{Adler}},  
\bibinfo{Title}{Photon splitting and photon dispersion in a strong magnetic field},
\bibinfo{journal}{Annals Phys.} \pmb{\bibinfo{volume}{67}},
\bibinfo{pages}{599} (\bibinfo{year}{1971}).

\bibitem{Baring}
\bibinfo{author}{\bibfnamefont{M.~G.}~\bibnamefont{Baring}}, 
\bibinfo{Title}{Title: Photon-splitting limits to the hardness of emission in strongly magnetized soft gamma repeaters},
\bibinfo{journal}{Astrophys. J. Lett.} \pmb{\bibinfo{volume}{440}},
\bibinfo{pages}{L69}  (\bibinfo{year}{1995}).

\bibitem{Baring:2000cr}
\bibinfo{author}{\bibfnamefont{M.~G.}~\bibnamefont{Baring}} 
\bibnamefont{and}
\bibinfo{author}{\bibfnamefont{A.~K.}~\bibnamefont{Harding}}, 
\bibinfo{Title}{Photon splitting and pair creation in highly magnetized pulsars},
\bibinfo{journal}{Astrophys. J.} \pmb{\bibinfo{volume}{547}},
\bibinfo{pages}{929}  (\bibinfo{year}{2001}).

\bibitem{Usov:2002df}
\bibinfo{author}{\bibfnamefont{V.~V.}~\bibnamefont{Usov}}, 
\bibinfo{Title}{Photon splitting in the superstrong magnetic fields of pulsars},
\bibinfo{journal}{Astrophys. J. Lett.} \pmb{\bibinfo{volume}{572}},
\bibinfo{pages}{L87}  (\bibinfo{year}{2002}).

\bibitem{Chistyakov:2012ms}
\bibinfo{author}{\bibfnamefont{M.~V.}~\bibnamefont{Chistyakov}}, 
\bibinfo{author}{\bibfnamefont{D.~A.}~\bibnamefont{Rumyantsev}}, \bibnamefont{and}
\bibinfo{author}{\bibfnamefont{N.~S.}~\bibnamefont{Stus'}}, 
\bibinfo{Title}{Photon splitting and Compton scattering in strongly magnetized hot plasma},
\bibinfo{journal}{Phys. Rev. D} \pmb{\bibinfo{volume}{86}},
\bibinfo{pages}{043007}  (\bibinfo{year}{2012}).

\bibitem{ShabadNature}
\bibinfo{author}{\bibfnamefont{A.~E.}~\bibnamefont{Shabad}} 
\bibnamefont{and}
\bibinfo{author}{\bibfnamefont{V.~V.}~\bibnamefont{Usov}}, 
\bibinfo{Title}{$\gamma$-Quanta capture by magnetic field and pair creation suppression in pulsars},
\bibinfo{journal}{Nature} \pmb{\bibinfo{volume}{295}},
\bibinfo{pages}{215} (\bibinfo{year}{1982}).

\bibitem{ShabadAstroSpaSci}
\bibinfo{author}{\bibfnamefont{A.~E.}~\bibnamefont{Shabad}} 
\bibnamefont{and}
\bibinfo{author}{\bibfnamefont{V.~V.}~\bibnamefont{Usov}}, 
\bibinfo{Title}{Propagation of $\gamma$-radiation in strong magnetic fields of pulsars},
\bibinfo{journal}{Astrophys. Space Sci.} \pmb{\bibinfo{volume}{102}},
\bibinfo{pages}{327} (\bibinfo{year}{1984}).

\bibitem{Herold}
\bibinfo{author}{\bibfnamefont{H.}~\bibnamefont{Herold}}, 
\bibinfo{author}{\bibfnamefont{H.}~\bibnamefont{Ruder}}, 
\bibnamefont{and}
\bibinfo{author}{\bibfnamefont{G.}~\bibnamefont{Wunner}}, 
\bibinfo{Title}{Can quanta really be captured by pulsar magnetic fields?},
\bibinfo{journal}{Phys. Rev. Lett.} \pmb{\bibinfo{volume}{54}},
\bibinfo{pages}{1452} (\bibinfo{year}{1985}).

\bibitem{Bathia}
\bibinfo{author}{\bibfnamefont{V.~B.}~\bibnamefont{Bathia}},
\bibinfo{author}{\bibfnamefont{N.}~\bibnamefont{Chopra}}, 
\bibnamefont{and}
\bibinfo{author}{\bibfnamefont{N.}~\bibnamefont{Panchapakesan}}, 
\bibinfo{Title}{Photon capture in pulsar magnetic fields},
\bibinfo{journal}{Astrophys. Space Sci.} \pmb{\bibinfo{volume}{129}},
\bibinfo{pages}{271} (\bibinfo{year}{1987}).

\bibitem{Selym2}
\bibinfo{author}{\bibfnamefont{S.}~\bibnamefont{Villalba-Ch\'avez}}, \bibinfo{author}{\bibfnamefont{A.~E.}~\bibnamefont{Shabad}} \bibnamefont{and}
\bibinfo{author}{\bibfnamefont{C.}~\bibnamefont{M\"uller}}, 
\bibinfo{Title}{Magnetic dominance of axion electrodynamics: Photon capture effect and anisotropy of Coulomb potential},
\bibinfo{journal}{Eur. Phys. J. C} \pmb{\bibinfo{volume}{81}},
\bibinfo{pages}{331} (\bibinfo{year}{2021}).

\bibitem{Manchester} 
\bibinfo{author}{\bibfnamefont{R.}~\bibfnamefont{M.}~\bibnamefont{Manchester}},
\bibinfo{author}{\bibfnamefont{G.~B.}~\bibnamefont{Hobbs}}, \bibinfo{author}{\bibfnamefont{A.}~\bibnamefont{Teoh}}
and  \bibinfo{author}{\bibfnamefont{M.}~\bibnamefont{Hobbs}},  
\bibinfo{title}{The Australia Telescope National Facility Pulsar Catalogue},
\bibinfo{journal}{Astron. J.} \textbf{\bibinfo{volume}{129}}, 
\bibinfo{pages}{1993} (\bibinfo{year}{2005}). 

\bibitem{Kouveliotou}
\bibinfo{author}{\bibfnamefont{C.}~\bibnamefont{Kouveliotou}}  \bibnamefont{\emph{et al.}}, 
\bibinfo{Title}{An X-ray pulsar with a superstrong magnetic field in the soft gamma-ray repeater SGR 1806-20},
\bibinfo{journal}{Nature} \pmb{\bibinfo{volume}{393}},
\bibinfo{pages}{235} (\bibinfo{year}{1998}).

\bibitem{Alaa}
\bibinfo{author}{\bibfnamefont{A.~I.}~\bibnamefont{Ibrahim}}  \bibnamefont{\emph{et al.}},  
\bibinfo{tilte}{New evidence for proton cyclotron resonance in a magnetar strength field from SGR 1806-20}, 
\bibinfo{journal}{Astrophys. J. Lett.} \textbf{\bibinfo{volume}{584}},
\bibinfo{pages}{L17} (\bibinfo{year}{2003}).

\bibitem{Gnedin}
\bibinfo{author}{\bibfnamefont{Y.~N.}~\bibnamefont{Gnedin} \emph{et al.}},  
\bibinfo{tilte}{Radio emission of the magnetar SGR 1806-20: Evolution of the magnetic field in the region of the radio afterglow},
\bibinfo{journal}{Astron. Rep.} \textbf{\bibinfo{volume}{51}},
\bibinfo{pages}{863} (\bibinfo{year}{2007})

\bibitem{Ibrahim}
\bibinfo{author}{\bibfnamefont{A.~I.}~\bibnamefont{Ibrahim}}  \bibnamefont{\emph{et al.}}, 
\bibinfo{tilte}{Discovery of a Transient Magnetar: XTE J1810-197}, 
\bibinfo{journal}{Astrophys. J.} \textbf{\bibinfo{volume}{609}},
\bibinfo{pages}{L17} (\bibinfo{year}{2004}).

\bibitem{McGill}
\bibinfo{author}{\bibfnamefont{S.~A.}~\bibnamefont{Olausen}}   \bibnamefont{and} 
\bibinfo{author}{\bibfnamefont{V.~M.}~\bibnamefont{Kaspi}}, 
\bibinfo{tilte}{The McGill magnetars catalog}, 
\bibinfo{journal}{Astrophys. J. Supp.} \textbf{\bibinfo{volume}{212}},
\bibinfo{pages}{6} (\bibinfo{year}{2014}); www.physics.mcgill.ca/$\sim$pulsar/magnetar/main.html.

\bibitem{Kaspi}
\bibinfo{author}{\bibfnamefont{V.~M.}~\bibnamefont{Kaspi}}   \bibnamefont{and} 
\bibinfo{author}{\bibfnamefont{A.~M.}~\bibnamefont{Beloborodov}}, 
\bibinfo{tilte}{Magnetars}, 
\bibinfo{journal}{Annu. Rev. Astron. Astrophys.} \textbf{\bibinfo{volume}{55}},
\bibinfo{pages}{261} (\bibinfo{year}{2017}).

\bibitem{Kharzeev}
\bibinfo{author}{\bibfnamefont{D.~E.}~\bibnamefont{Kharzeev}}, 
\bibinfo{author}{\bibfnamefont{L.~D.}~\bibnamefont{McLerran}},  
 \bibnamefont{and} 
\bibinfo{author}{\bibfnamefont{H.~J.}~\bibnamefont{Warringa}}, 
\bibinfo{Title}{The effects of topological charge change in heavy ion collisions: Event by event P and CP violation},
\bibinfo{journal}{Nucl. Phys.  A} \pmb{\bibinfo{volume}{803}},
\bibinfo{pages}{227} (\bibinfo{year}{2008}).

\bibitem{Skokov}
\bibinfo{author}{\bibfnamefont{V.}~\bibnamefont{Skokov}}, 
\bibinfo{author}{\bibfnamefont{A.~Y.}~\bibnamefont{Illarionov}},  
 \bibnamefont{and} 
\bibinfo{author}{\bibfnamefont{V.}~\bibnamefont{Toneev}}, 
\bibinfo{Title}{Estimate of the magnetic field strength in heavy-ion collisions},
\bibinfo{journal}{Int. J. Mod. Phys. A} \pmb{\bibinfo{volume}{24}},
\bibinfo{pages}{5925} (\bibinfo{year}{2009}).

\bibitem{Zhong}
\bibinfo{author}{\bibfnamefont{Y.}~\bibnamefont{Zhong}}, 
\bibinfo{author}{\bibfnamefont{C.-B.}~\bibnamefont{Yang}},
\bibinfo{author}{\bibfnamefont{X.}~\bibnamefont{Cai}},  
 \bibnamefont{and} 
\bibinfo{author}{\bibfnamefont{S.-Q.}~\bibnamefont{Feng}}, 
\bibinfo{Title}{A systematic study of magnetic field in relativistic heavy-ion collisions in the RHIC and LHC energy regions},
\bibinfo{journal}{Adv. High Energy Phys.} \pmb{\bibinfo{volume}{2014}},
\bibinfo{pages}{193039} (\bibinfo{year}{2014}).

\bibitem{Brandenburg}
\bibinfo{author}{\bibfnamefont{J.~D.}~\bibnamefont{Brandenburg}}, 
\bibinfo{author}{\bibfnamefont{W.}~\bibnamefont{Zha}},  
 \bibnamefont{and} 
\bibinfo{author}{\bibfnamefont{Z.}~\bibnamefont{Xu}}, 
\bibinfo{Title}{Mapping the electromagnetic fields of heavy-ion collisions with Breit-Wheeler process},
\bibinfo{journal}{Eur. Phys. J.  A} \pmb{\bibinfo{volume}{57}},
\bibinfo{pages}{299} (\bibinfo{year}{2021}).

\bibitem{Vashaspati}
\bibinfo{author}{\bibfnamefont{T.}~\bibnamefont{Vashaspati}}, 
\bibinfo{Title}{Magnetic field from cosmological phase transitions},
\bibinfo{journal}{Phys. Lett. B} \pmb{\bibinfo{volume}{265}},
\bibinfo{pages}{258} (\bibinfo{year}{1991}).

\bibitem{Vachaspati:2020blt}
\bibinfo{author}{\bibfnamefont{T.}~\bibnamefont{Vashaspati}}, 
\bibinfo{Title}{Progress on cosmological magnetic fields,},
\bibinfo{journal}{Rept. Prog. Phys.} \pmb{\bibinfo{volume}{84}},
\bibinfo{pages}{074901} (\bibinfo{year}{2021}).

\bibitem{Enqvist}
\bibinfo{author}{\bibfnamefont{K.}~\bibnamefont{Enqvist}}, 
\bibinfo{author}{\bibfnamefont{P.}~\bibnamefont{Olensen}},  \bibnamefont{and} 
\bibinfo{author}{\bibfnamefont{V.}~\bibnamefont{Semikoz}}, 
\bibinfo{Title}{Galactic dynamo and nucleosynthesis limit on the dirac Neutrino Masses},
\bibinfo{journal}{Phys. Rev. Lett.} \pmb{\bibinfo{volume}{69}},
\bibinfo{pages}{2157} (\bibinfo{year}{1992}).

\bibitem{Baym}
\bibinfo{author}{\bibfnamefont{G.}~\bibnamefont{Baym}}, 
\bibinfo{author}{\bibfnamefont{D.}~\bibnamefont{B\"odeker}},  
 \bibnamefont{and} 
\bibinfo{author}{\bibfnamefont{L.}~\bibnamefont{McLerran}}, 
\bibinfo{Title}{Magnetic fields produced by phase transition bubbles in the electroweak phase transition},
\bibinfo{journal}{Phys. Rev. D} \pmb{\bibinfo{volume}{53}},
\bibinfo{pages}{662} (\bibinfo{year}{1996}).

\bibitem{Grasso:2000wj}
\bibinfo{author}{\bibfnamefont{D.}~\bibnamefont{Grasso}},   
 \bibnamefont{and} 
\bibinfo{author}{\bibfnamefont{H.~R.}~\bibnamefont{Rubinstein}}, 
\bibinfo{Title}{Magnetic fields in the early universe},
\bibinfo{journal}{Phys. Rept.} \pmb{\bibinfo{volume}{348}},
\bibinfo{pages}{163} (\bibinfo{year}{2001}).

%%%%%%%%%%%%%%%%%%%%%THEORETICAL FRAMEWORK%%%%%%%%%%%%%%%%%%%%%%%

\bibitem{Fradkin}
\bibinfo{author}{\bibfnamefont{E.~S.}~\bibnamefont{Fradkin}},
\bibinfo{Title}{Quantum field theory and hydrodynamics},
\bibinfo{journal}{Proceeding (Trudy) of the P. N. Lebedev Phys. Inst.} \pmb{\bibinfo{volume}{29}},
\bibinfo{journal}{Consultants Bureau, New York},  (\bibinfo{year}{1967}).

\bibitem{Holger}
\bibinfo{author}{\bibfnamefont{H.}~\bibnamefont{Gies}} \bibnamefont{and}
\bibinfo{author}{\bibfnamefont{W.}~\bibnamefont{Dittrich}}, 
\bibinfo{Title}{Probing the quantum vacuum: Perturbative effective action approach in quantum electrodynamics and its application},
\bibinfo{journal}{Springer Heidelberg, Band 166}, (\bibinfo{year}{2000}).

\bibitem{Weinberg}
\bibinfo{author}{\bibfnamefont{S.}~\bibnamefont{Weinberg}}, 
\bibinfo{Title}{The quantum theory of the field, Vol I},
\bibinfo{journal}{Cambridge Uni. Press}, (\bibinfo{year}{1995}).

\bibitem{Landau}
\bibinfo{author}{\bibfnamefont{L.~D.}~\bibnamefont{Landau}} \bibnamefont{and}
\bibinfo{author}{\bibfnamefont{E.~M.}~\bibnamefont{Lifshitz}}, 
\bibinfo{Title}{Electrodynamics of continuous media},
\bibinfo{Title}{Course of theoretical physics; Vol. 8},
\bibinfo{journal}{Pergamon Press, Oxford},
\bibinfo{pages}{Chap. 9} (\bibinfo{year}{1984}).

%%%%%%%%%%%%%%%THERMAL FIELD THEORY%%%%%%%%%%%%%%%%%%%%%%%%%%%%%%%

\bibitem{Bordag}
\bibinfo{author}{\bibfnamefont{M.}~\bibnamefont{Bordag}}, \bibinfo{author}{\bibfnamefont{K.}~\bibnamefont{Kirsten}} \bibnamefont{and}
\bibinfo{author}{\bibfnamefont{D.~V.}~\bibnamefont{Vassilevich}}, 
\bibinfo{Title}{Path-integral quantization of electrodynamics in dielectric media},
\bibinfo{journal}{J. Phys. A} \pmb{\bibinfo{volume}{31}},
\bibinfo{pages}{2381} (\bibinfo{year}{1998}).

\bibitem{mmph}
\bibinfo{author}{\bibfnamefont{H.~P\'erez.}~\bibnamefont{Rojas}}  \bibnamefont{and}
\bibinfo{author}{\bibfnamefont{E.~Rodriguez}~\bibnamefont{Querts}}, 
\bibinfo{Title}{Is the photon paramagnetic?},
\bibinfo{journal}{Phys. Rev. D} \pmb{\bibinfo{volume}{79}},
\bibinfo{pages}{093002} (\bibinfo{year}{2009}).

\bibitem{mmps}
\bibinfo{author}{\bibfnamefont{S.~}~\bibnamefont{Villalba-Ch\'avez}},  
\bibinfo{title}{Photon magnetic moment and vacuum magnetization in an asymptotically large magnetic field}, 
\bibinfo{journal}{Phys. Rev. D} \textbf{\bibinfo{volume}{81}}, \bibinfo{pages}{105019}  (\bibinfo{year}{2010}).

\bibitem{Villalba-Chavez:2012pmx}
\bibinfo{author}{\bibfnamefont{S.~}\bibnamefont{Villalba-Ch\'avez}} ~\bibnamefont{and}~ 
\bibinfo{author}{\bibfnamefont{A.~E.~}\bibnamefont{Shabad}},
 \bibinfo{title}{QED with external field: Hamiltonian treatment for anisotropic medium formed by the Lorentz-non-invariant vacuum},  \bibinfo{journal}{Phys. Rev. D}  \pmb{\bibinfo{volume}{86}}, 
\bibinfo{pages}{105040}  (\bibinfo{year}{2012}).


%%%%%%%%%%%%%%%%%%% RESULTS %%%%%%%%%%%%%%%%%%%%%%%%%%%%%%%%%%

\bibitem{Mercier}
\bibinfo{author}{\bibfnamefont{R.~P.}~\bibnamefont{Mercier}},  
\bibinfo{title}{Thermal radiation in anisotropic media}, 
\bibinfo{journal}{Proc. Phys. Soc.} \textbf{\bibinfo{volume}{83}}, \bibinfo{pages}{811}  (\bibinfo{year}{1963}).

\bibitem{Cole}
\bibinfo{author}{\bibfnamefont{K.~D.}~\bibnamefont{Cole}},  
\bibinfo{title}{Generalization of Plank's law of radiation to  anisotropic  dispersive media}, 
\bibinfo{journal}{Aust. J. Phys.} \textbf{\bibinfo{volume}{30}}, \bibinfo{pages}{671}  (\bibinfo{year}{1977}).

\bibitem{Eric}
\bibinfo{author}{\bibfnamefont{E.}~\bibnamefont{Kochems}},  
\bibinfo{author}{\bibfnamefont{G.}~\bibnamefont{Quintero} \bibnamefont{Angulo}},  
\bibinfo{author}{\bibfnamefont{R.}~\bibnamefont{Egger}}, 
\bibinfo{author}{\bibfnamefont{C.}~\bibnamefont{M\"uller}},~\bibnamefont{and}~\bibinfo{author}{\bibfnamefont{S.}~\bibnamefont{Villalba-Ch\'avez}},  
\bibinfo{title}{Magnetic-field-tunable anisotropic blackbody radiation and condensation of slow thermal light in dynamical axion insulators}, 
\bibinfo{journal}{Phys. Rev. Res.}  \textbf{\bibinfo{volume}{7}}, \bibinfo{pages}{033297}  (\bibinfo{year}{2025}).

\bibitem{Ferrer} 
\bibinfo{author}{\bibfnamefont{E.~J.~}\bibnamefont{Ferrer}}~\bibnamefont{and}~\bibinfo{author}{\bibfnamefont{A.}\bibnamefont{Sanchez}}, 
\bibinfo{title}{Magnetic field effect in the fine-structure constant and electron dynamical mass},
\bibinfo{journal}{Phys. Rev. D}  \pmb{\bibinfo{volume}{100}}, 
\bibinfo{pages}{096006}  (\bibinfo{year}{2019}).

\bibitem{ShabadLangrangian} 
\bibinfo{author}{\bibfnamefont{A.~E.~}\bibnamefont{Shabad}}~\bibnamefont{and}~\bibinfo{author}{\bibfnamefont{V.~V.}\bibnamefont{Usov}}, 
\bibinfo{title}{Effective Lagrangian in nonlinear electrodynamics and its properties of causality and unitarity},
\bibinfo{journal}{Phys. Rev. D}  \pmb{\bibinfo{volume}{83}}, 
\bibinfo{pages}{105006}  (\bibinfo{year}{2011}).

\bibitem{vanHove} 
\bibinfo{author}{\bibfnamefont{L.~}\bibnamefont{van Hove}}, 
\bibinfo{title}{The occurrence of singularities in the elastic frequency distribution of a crystal},
\bibinfo{journal}{Phys. Rev.}  \pmb{\bibinfo{volume}{89}}, 
\bibinfo{pages}{1189}  (\bibinfo{year}{1953}).
 
\bibitem{Chamel}
\bibinfo{author}{\bibfnamefont{N.}~\bibnamefont{Chamel}}~\bibnamefont{and}~
\bibinfo{author}{\bibfnamefont{P.}~\bibnamefont{Haensel}}, 
\bibinfo{Title}{Physics of Neutron Star Crusts}, 
\bibinfo{journal}{Living Rev. Relativity}  \pmb{\bibinfo{volume}{11}}, 
\bibinfo{pages}{10} (\bibinfo{year}{2008}).
 
 \bibitem{shapiro}
\bibinfo{author}{\bibfnamefont{S.~L.}~\bibnamefont{Shapiro}}~\bibnamefont{and}~\bibinfo{author}{\bibfnamefont{S.~A.}\bibnamefont{Teukolsky}}, 
\bibinfo{Title}{Black holes, white dwarfs, and neutron stars: The physics of compact objects}, 
 \bibinfo{journal}{Wiley-VCH Verlag  GmbH \&  Co. KGaA  Weinheim},  (\bibinfo{year}{2004}).
 
 \bibitem{Prakash}
\bibinfo{author}{\bibfnamefont{J.~M.}~\bibnamefont{Lattimer}}~\bibnamefont{and}~
\bibinfo{author}{\bibfnamefont{M.}~\bibnamefont{Prakash}}, 
\bibinfo{Title}{The equation of state of hot, dense matter and neutron stars}, 
 \bibinfo{journal}{Phys. Rep.}  \pmb{\bibinfo{volume}{621}}, 
\bibinfo{pages}{127} (\bibinfo{year}{2016}).
 
 \bibitem{Haensel}
\bibinfo{author}{\bibfnamefont{P.}~\bibnamefont{Haensel}}, 
\bibinfo{author}{\bibfnamefont{A.~Y.}~\bibnamefont{Potekhin}}, 
~\bibnamefont{and}~\bibinfo{author}{\bibfnamefont{D.~G.}\bibnamefont{Yakovlev}}, 
\bibinfo{Title}{Neutron Stars 1: Equation of State and Structure}, 
 \bibinfo{journal}{Springer New York},  (\bibinfo{year}{2007}).

 \bibitem{PRLPrakash}
\bibinfo{author}{\bibfnamefont{J.~M.}~\bibnamefont{Lattimer}}~\bibnamefont{and}~
\bibinfo{author}{\bibfnamefont{M.}~\bibnamefont{Prakash}}, 
\bibinfo{Title}{Ultimate energy density of observable cold baryonic matter}, 
 \bibinfo{journal}{Phys. Rev. Lett.}  \pmb{\bibinfo{volume}{94}}, 
\bibinfo{pages}{111101} (\bibinfo{year}{2005}).
  
\end{thebibliography}
\end{document}